\documentclass[aps,prl,showpacs,twocolumn,floats]{revtex4}
\usepackage{amssymb}

\usepackage{graphicx,psfig}
\usepackage{dcolumn}
\usepackage{bm}

\input{tcilatex}

\begin{document}

\title{Quantum Nonlinear Switching Model \vspace{-1mm} }
\author{D. A. Garanin and R Schilling}
\affiliation{Institut f\"{u}r Physik, Johannes
Gutenberg-Universit\"{a}t, D-55099 Mainz, Germany}
\date{\today}

\begin{abstract}
We present a method, the dynamical cumulant expansion, that allows to
calculate quantum corrections for time-dependent quantities of interacting
spin systems or single spins with anisotropy. This method is applied to the
quantum-spin model $\widehat{H}=-H_{z}(t)S_{z}+V(\mathbf{S})$ with $%
H_{z}(\pm \infty )=\pm \infty $ and $\Psi (-\infty )=\left| -S\right\rangle $
we study the quantity $P(t)=\left( 1-\left\langle S_{z}\right\rangle
_{t}/S\right) /2.$ The case $V(\mathbf{S})=-H_{x}S_{x}$ corresponds to the
standard Landau-Zener-Stueckelberg model of tunneling at avoided-level
crossing for $N=2S$ independent particles mapped onto a single-spin-$S$
problem, $P(t)$ being the staying probability. Here the solution does not
depend on $S$ and follows, e.g., from the classical Landau-Lifshitz
equation. A term $-DS_{z}^{2}$ accounts for particles' interaction and it
makes the model nonlinear and essentially quantum mechanical. The $1/S$
corrections obtained with our method are in a good accord with a full
quantum-mechanical solution if the classical motion is regular, as for $D>0$.
\end{abstract}

\pacs{ 03.65.-w, 75.10.Jm, 05.45.Mt} \maketitle

The classical limit of quantum spin systems has been discussed for decades
in terms of spin coherent states \cite
{arecougiltho72,lie73,kla79,kursuz80,sch95}. Yet little is known, in
particular, about quantum corrections to the classical dynamics of systems
with large spin $S.$ One possible reason is that initially narrow quantum
wave packets are spreading with time $t$, i.e., quantum states depend on the
combined parameter of type $t^{\alpha }/S$ ($\alpha \geq 1$). Therefore the
limits $t\rightarrow \infty $ and $S\rightarrow \infty $ do not commute and
a perturbation theory in pure $1/S$ is usually impossible. One is left here
with the question of how well defined asymptotic ($t\rightarrow \infty $)
quantities behave in the limit $S\rightarrow \infty .$ Are they continuous
functions of $1/S$ for $1/S\rightarrow 0?$

One of the problems of this kind is a recent generalization \cite
{hamraemiysai00,gar03prb} of the Landau-Zener-Stueckelberg (LZS) problem
\cite{lan32,zen32,stu32} of transitions at avoided level crossings that is a
well known quantum effect ubiquitous in physics of atomic and molecular
collisions. Still the standard LZS effect formally allows classical
description since the Schr\"{o}dinger equation for a pseudospin $S=1/2$ is
equivalent to the classical Landau-Lifshitz (LL) equation
\begin{equation}
\mathbf{\dot{m}}=\gamma \left[ \mathbf{m\times H}_{\mathrm{eff}}\right]
,\qquad \mathbf{H}_{\mathrm{eff}}=-\partial \mathcal{H}/\partial \mathbf{m},
\label{LLE}
\end{equation}
where $\mathbf{m\equiv }\left\langle \mathbf{S}\right\rangle \mathbf{/}S,$ $%
\gamma =g\mu _{B}/\hbar $, and, in particular, $\mathbf{H}_{\mathrm{eff}}=%
\mathbf{H}=H_{x}\mathbf{e}_{x}+H_{z}(t)\mathbf{e}_{z}$ with $H_{x}=\Delta $
(level splitting) and $H_{z}(t)=E_{-1}(t)-E_{1}(t)$ (level bias) satisfying $%
H_{z}(\pm \infty )=\pm \infty .$ Typical time dependence of $H_{z}$ is
linear, $H_{z}=vt,$ or nonlinear \cite{garsch02prb}. The probability to stay
at the initially populated level $-1$ is
\begin{equation}
P(t)=\left[ 1-m_{z}(t)\right] /2,\qquad P(-\infty )=1  \label{PDef}
\end{equation}
and for the linear sweep \cite{lan32,zen32},
\begin{equation}
P(\infty )\equiv P=e^{-\varepsilon },\qquad \varepsilon \equiv \frac{\pi
\Delta ^{2}}{2\hbar v}.  \label{PLZ}
\end{equation}

This ``degeneracy'' between quantum and classical descriptions of the LZS
effect disappears if tunneling particles interact with each other, as is the
case for molecular magnets \cite{weretal00epl} (see Ref. \cite{sesgat03}\
for a recent review) or for the Bose-Einstein condensate \cite
{zobgar00,wuniu00,liuetal02}. Finding the time-dependent wave function for
the whole system of $N\gg 1$ particles that is specified by $2^{N}$
coefficients is of course a tremendous problem. As a plausible first step
one can consider a simplified model in which each particle interacts with
all $N-1$ other particles with the same strength $J$ \cite
{hamraemiysai00,gar03prb} that maps onto a large-spin model \cite{gar03prb}
(we use $g\mu _{B}=\hbar =1$ below)
\begin{eqnarray}
\widehat{H} &=&-H_{z}(t)S_{z}+V(\mathbf{S}),\qquad S=N/2  \label{HanSDef} \\
V(\mathbf{S}) &=&-H_{x}S_{x}-DS_{z}^{2},\qquad D=2J,  \label{VSAxial}
\end{eqnarray}
whereas $P(t)$ is still defined by Eq. (\ref{PDef}). On one hand, our
original model of interacting particles \cite{hamraemiysai00,gar03prb} in
the limit $N\rightarrow \infty $ is described exactly by the mean-field
approximation (MFA) that simplifies the problem to a \emph{nonlinear}
Schr\"{o}dinger equation for a single tunneling particle \cite
{zobgar00,wuniu00,liuetal02,hamraemiysai00}. On the other hand, the
mean-field limit $N\rightarrow \infty $ corresponds to the classical limit $%
S\rightarrow \infty $ of Eq. (\ref{HanSDef}) that results into the LL
equation Eq. (\ref{LLE}) with classical energy $\mathcal{H}=-H_{z}(t)m_{z}-v(%
\mathbf{m}),$ where [not to be confused with the sweep rate $v$] $v(\mathbf{m%
})=-H_{x}m_{x}-dm_{z}^{2}$ and $d\equiv SD.$ However for $N\neq \infty $
that better suits to real systems with finite range interactions one is
confronted with an \emph{essentially} \emph{quantum effect}.

One can also look upon the large-spin problem of Eq. (\ref{HanSDef}) from
another perspective and consider it as a kind of scattering problem for a
spin: Sweeping the field across the region of strong interaction, $H_{z}(\pm
\infty )=\pm \infty ,$ with the task to find $\left\langle
S_{z}\right\rangle _{t\rightarrow \infty }$, if in the initial state the
spin was down, $\left\langle S_{z}\right\rangle _{t=-\infty }=-S.$ We call
it the ``Quantum Nonlinear Switching (QNS) model''. This model is more
general than the LZS model and it can be applied to \emph{real} large spins
rather than to large pseudospins composed of many pseudospins 1/2. It could
be relevant for molecular magnets, and one can even include in $V(\mathbf{S})
$ terms that are absent in the LZS\ tunneling model, e.g., the biaxial
anisotropy $E(S_{x}^{2}-S_{y}^{2})$ characteristic to Fe$_{8}$.

The purpose of this Letter is to investigate the semiclassical limit $S\gg 1$
and to work out an analytical method to calculate $1/S$ corrections to the
mean-field result for the staying probability $P$ for the QNS model. This
provides a crucial test for the MFA that was applied to the LZS effect in
systems of interacting particles in the absence of a more accurate and still
tractable method\cite{hamraemiysai00,gar03prb}. In comparison to statics,
application of the MFA to dynamics is questionable since the stability of
the solution of the nonlinear LL equation Eq. (\ref{LLE}) with respect to
quantum fluctuations is not guaranteed. It becomes clear as soon as one
realizes that dynamical MFA describes \emph{classical} dynamics that might
exhibit chaos, that is, classically computed quantities $P(t)$ might be
poorly defined.

\begin{figure}[t]
\unitlength1cm
\begin{picture}(11,6)
\centerline{\psfig{file=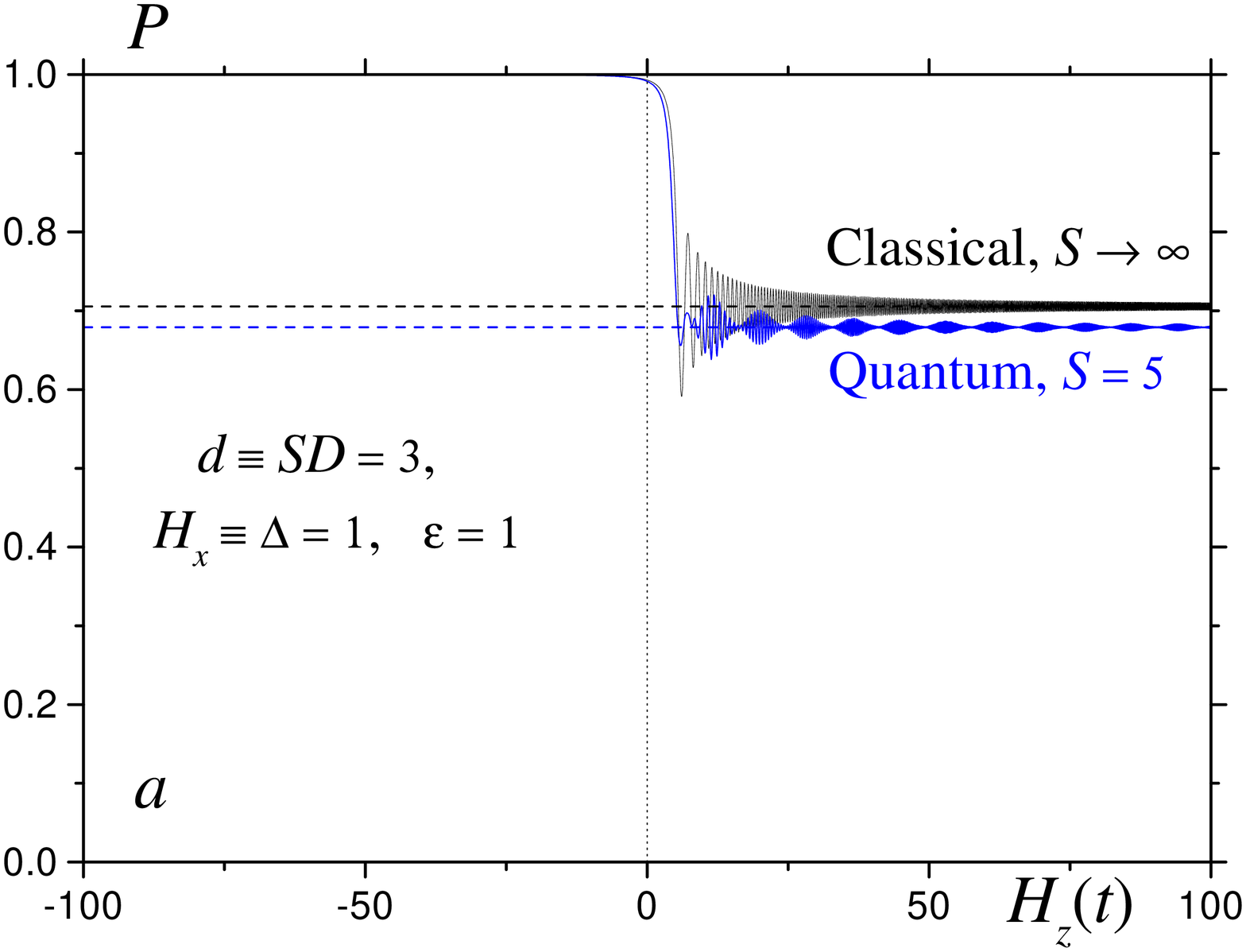,angle=-90,width=9cm}}
\end{picture}
\begin{picture}(11,6)
\centerline{\psfig{file=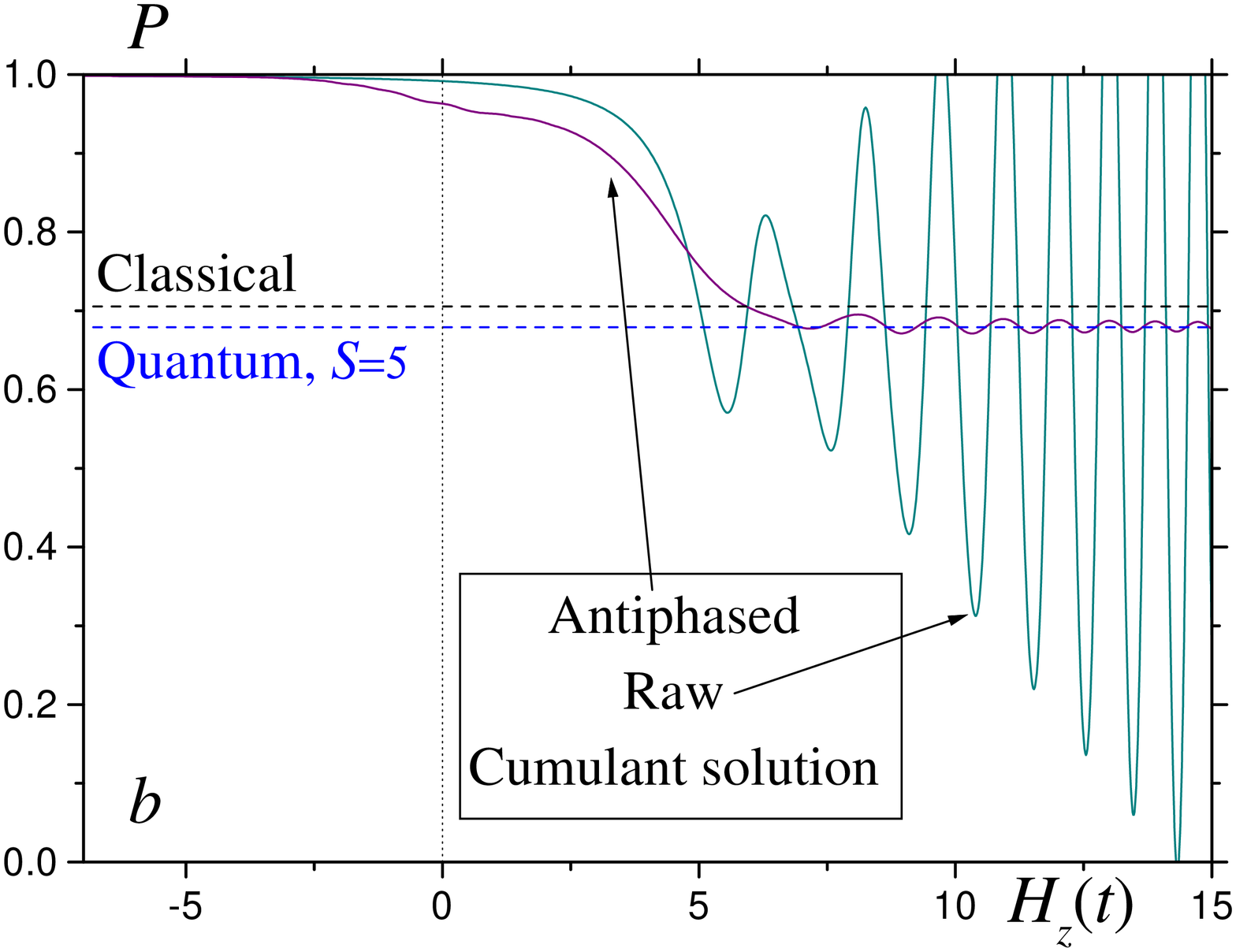,angle=-90,width=9cm}}
\end{picture}
\caption{ \label{Fig-LZQ-PvsHz}
Staying probability $P(t)$ for the quantum nonlinear LZS effect.
$a$ -- Exact numerical solutions for $S=5$ and $S\to\infty$;
$b$ -- Cumulant solution of order $1/S$: Raw and antiphased (see text).
}
\end{figure}

The method we propose here is a dynamic cumulant expansion (DCE) based on
the normalized and symmetrized spin cumulants that are defined by
\begin{eqnarray}
m_{\alpha } &\equiv &\left\langle S_{\alpha }\right\rangle /S,  \nonumber \\
m_{\alpha \beta } &\equiv &\frac{1}{S^{2}}\left( \frac{\left\langle
S_{\alpha }S_{\beta }\right\rangle +\left\langle S_{\beta }S_{\alpha
}\right\rangle }{2}-\left\langle S_{\alpha }\right\rangle \left\langle
S_{\beta }\right\rangle \right)   \label{CumsDef}
\end{eqnarray}
etc. with $\alpha ,\beta =x,y,z=1,2,3.$ We use the Heisenberg representation
for spin operators $S_{\alpha }$ and compute matrix elements with respect to
the initial state that is the particular case $\theta =\pi $ of the spin
coherent state  $\left| \Omega (\theta ,\varphi )\right\rangle
=\sum_{m=-S}^{S}C_{m}\left| m\right\rangle $ with
\begin{equation}
C_{m}=\binom{2S}{S+m}^{1/2}\left( \cos \frac{\theta }{2}\right) ^{S+m}\left(
\sin \frac{\theta }{2}\right) ^{S-m}e^{i(S-m)\varphi }  \label{Coherent}
\end{equation}
\cite{arecougiltho72,lie73,kla79,kursuz80}. Matrix elements of products of
operators factorize in the classical limit, and, correspondingly, normalized
cumulants of $n$ spin operators are as small as $S^{1-n}$\cite
{klafulgar99,garklaful00}$.$ Thus a theory formulated in terms of cumulants
yields a quasiclassical expansion. It is sufficient to retain pair cumulants
and drop higher-order cumulants to obtain $1/S$ corrections. If the state of
the system would be close to a coherent state during all the time, the
cumulant method would yield the result without any difficulties. We will
see, however, that this assumption fails at large times because the
nonequidistant spectrum of a nonlinear spin system leads to dephasing of
different $\left| m\right\rangle $-components of the coherent state. One can
think of a strongly localized cloud of classical spins that, however, are
precessing with slightly different frequencies. Whatever narrow is this
cloud, it will spread along the classical trajectory within a dephasing time
$T_{\mathrm{deph}}$. That is, cumulants cease to be small at large times and
the DCE diverges. Mathematically it means, as we will see below, that $P(t)$
depends on both $S^{-1}$ and on the combined parameter $S^{-1}t^{2},$ where
the pure $S^{-1}$ describes the quantum correction to the QNS and $%
S^{-1}t^{2}$ describes the dephasing effect. The DCE picks both
terms and thus it diverges at $t\to\infty$. Fortunately this
divergence can be eliminated to yield a pure $1/S$ correction to
$P\equiv P(\infty ),$ as shown in Fig. \ref{Fig-LZQ-PvsHz} for the
model described by Eq. (\ref {HanSDef}).

%

To construct DCE, we use the Heisenberg equation of motion for spin
operators with an interaction $V(\mathbf{S})$ that is taken to be a fully
symmetrized function of $S_{\alpha }$
\begin{eqnarray}
&&\dot{S}_{\alpha }=\epsilon _{\alpha \beta \gamma }\widehat{H}_{\gamma }^{%
\mathrm{eff}}(t,\mathbf{S})S_{\beta }  \label{SOperEq} \\
&&\widehat{H}_{\gamma }^{\mathrm{eff}}(t,\mathbf{S})\equiv H_{\gamma
}(t)-\partial V(\mathbf{S})/\partial S_{\gamma }.  \label{HeffDef}
\end{eqnarray}
Eq. (\ref{SOperEq}) is in fact fully symmetrized but we don't write it
explicitly to save space. Our quantity of interest is $m_{\alpha
}(t)=\left\langle -S\left| S_{\alpha }(t)\right| -S\right\rangle /S.$ Taking
this matrix element makes Eq. (\ref{SOperEq}) nonclosed because $\partial V(%
\mathbf{S})/\partial S_{\gamma }$ is an operator. After expanding the matrix
element of Eq. (\ref{SOperEq}) up to the pair cumulants and dropping
higher-order ones being of order $O(1/S^{2})$ it becomes an equation for $%
m_{\alpha }$ coupled to the pair cumulants $m_{\alpha \beta }.$ Equation for
the latter also follows from Eq. (\ref{SOperEq}) and it is closed at order $%
1/S.$ Initial conditions to these equations at $t=-\infty $ are $m_{z}=-1$
and $m_{xx}=m_{yy}=1/\left( 2S\right) $ while other components are zero. As
equations for $m_{\alpha }$ and $m_{\alpha \beta }$ are coupled, its
solution generates all powers of $1/S$ that should be suppressed to obtain a
pure $1/S$ expansion. To this end, we expand
\begin{equation}
m_{\alpha }=\mu _{\alpha }^{(0)}+\frac{1}{2S}\mu _{\alpha }^{(1)}+\ldots
,\qquad m_{\alpha \beta }=\frac{1}{2S}\mu _{\alpha \beta }^{(1)}+\ldots
\label{muDef}
\end{equation}
It is convenient to express the results in terms of
\begin{eqnarray}
v_{\alpha _{1}\cdots \alpha _{n}}(\mathbf{m}) &\equiv &S^{n-1}V_{\alpha
_{1}\cdots \alpha _{n}}(S\mathbf{m})  \label{vShortDef} \\
V_{\alpha _{1}\cdots \alpha _{n}}(\mathbf{S}) &\equiv &\frac{\partial ^{n}V(%
\mathbf{S})}{\partial S_{\alpha _{1}}\ldots \partial S_{\alpha _{n}}}.
\label{VShortDef}
\end{eqnarray}
As a result one obtains the three equations
\begin{eqnarray}
\dot{\mu}_{\alpha }^{(0)} &=&T_{\alpha \beta }(t,\mathbf{\mu }^{(0)})\mu
_{\beta }^{(0)}  \label{muEq0} \\
\dot{\mu}_{\alpha \beta }^{(1)} &=&\tilde{T}_{\alpha \gamma }(t,\mathbf{\mu }%
^{(0)})\mu _{\gamma \beta }^{(1)}+\tilde{T}_{\beta \gamma }(t,\mathbf{\mu }%
^{(0)})\mu _{\alpha \gamma }^{(1)}  \label{mumuEq1} \\
\dot{\mu}_{\alpha }^{(1)} &=&\tilde{T}_{\alpha \beta }(t,\mathbf{\mu }%
^{(0)})\mu _{\beta }^{(1)}+k_{\alpha \beta \gamma }(\mathbf{\mu }^{(0)})\mu
_{\beta \gamma }^{(1)},  \label{muEq1}
\end{eqnarray}
where
\begin{eqnarray}
T_{\alpha \beta }(t,\mathbf{m}) &=&\epsilon _{\alpha \beta \gamma }H_{\gamma
}^{\mathrm{eff}}(t,\mathbf{m})  \label{TTDef} \\
\tilde{T}_{\alpha \beta }(t,\mathbf{m}) &=&T_{\alpha \beta }(t,\mathbf{m}%
)-\epsilon _{\alpha \gamma \delta }m_{\gamma }v_{\delta \beta }(\mathbf{m})
\label{TTtilmDef}
\end{eqnarray}
with $H_{\gamma }^{\mathrm{eff}}(t,\mathbf{m})=H_{\gamma }(t)-v_{\gamma }(%
\mathbf{m})$ and
\begin{equation}
k_{\alpha \beta \gamma }(\mathbf{m})=-\epsilon _{\alpha \beta \delta
}v_{\delta \gamma }(\mathbf{m})-\epsilon _{\alpha \delta \eta }\frac{1}{2!}%
v_{\beta \gamma \eta }(\mathbf{m})m_{\delta }.  \label{kmDef}
\end{equation}
Equations (\ref{muEq0})--(\ref{muEq1}) should be solved numerically in order
of their appearance. The closed Eq. (\ref{muEq0}) is a \emph{nonlinear}
Landau-Lifshitz equation, whereas Eqs. (\ref{mumuEq1}) and (\ref{muEq1}) are
\emph{linear}. Although Eqs. (\ref{muEq0})--(\ref{muEq1}) and the method of
their derivation resemble decoupling schemes that are usually not based on
small parameters, the DCE is a rigorous (although formal) expansion in
powers of $1/S.$

Let us now check Eqs. (\ref{muEq0})--(\ref{muEq1}) for the exactly solvable
toy model of Eqs. (\ref{HanSDef}) with $H_{x}=0,$ starting at $t=0$ with an
arbitrarily directed spin coherent state, Eq. (\ref{Coherent}). Their
solution is nothing else than a $1/S$ expansion of the \emph{exact} solution
taken for
\begin{equation}
t\ll T_{\mathrm{rec}}=\pi /D=\pi S/d,  \label{TrecDef}
\end{equation}
where $T_{\mathrm{rec}}$ is the recurrence period, i.e., $1/S$ expansion of
\begin{equation}
m_{x}(t)\pm im_{y}(t)=e^{\mp i\left( \omega t-\varphi \right) }\exp \left[ -%
\frac{d^{2}t^{2}}{S}\sin ^{2}\theta \right] \sin \theta .
\label{mplusminusExactSmallt}
\end{equation}
In fact this expansion is only valid for $t\ll T_{\mathrm{deph}},$ where
\begin{equation}
T_{\mathrm{deph}}=\sqrt{S}/d\ll T_{\mathrm{rec}}  \label{TdephDef}
\end{equation}
is the dephasing time mentioned above Eq. (\ref{SOperEq}).

Consideration of the toy model shows that spin switching and
dephasing are different effects that are well separated in time
for $S\gg 1$: QNS occurs in the vicinity of the resonace,
$T_{QNS}\sim S^{0},$ while dephasing occurs much later,
$T_{\mathrm{deph}}\sim S^{1/2}$. Thus one concludes that the
divergence of DCE because of dephasing is harmless and it can be
cured. In the QNS setup, dephasing affects dynamics of $m_{z}(t)$
via $V(\mathbf{S}) $. For $S\rightarrow \infty $, time dependence
of $m_{z}$ far past the resonance can be found perturbatively in
$v(\mathbf{m})/H_{z}(t)\ll 1.$ Here precession of $\mathbf{m}$
around the $z$ axis is much faster than the temporal change of
$H_{z},$ i.e., $\dot{H}_{z}(t)/H_{z}^{2}(t)\ll 1.$ Thus the
classical energy $\mathcal{H}$ is nearly conserved over the period
of precession, and it yields
\begin{eqnarray}
m_{z}(t) &=&m_{z}^{(\mathrm{As})}+\delta m_{z}(t)  \label{DeltamzDef} \\
\delta m_{z}(t) &=&v(\mathbf{m}^{(\mathrm{As})}(t))/H_{z}(t),
\label{deltamzClass}
\end{eqnarray}
where $\mathbf{m}^{(\mathrm{As})}(t)$ corresponds to the asymptotic form $%
\mathcal{H}^{(\mathrm{As})}=-H_{z}(t)m_{z}.$ Precession of $\mathbf{m}^{(%
\mathrm{As})}(t)$ causes oscillatory behavior of $\delta m_{z}(t)$ and thus
of $P(t)$ that however vanishes in the limit $H_{z}(t)\rightarrow \infty $
(see Fig. \ref{Fig-LZQ-PvsHz}a). For finite $S\gg 1,$ dephasing of $\delta
m_{z}(t)$ follows from the first-order quantum-mechanical perturbation
theory
\begin{equation}
\delta m_{z}(t)\cong \frac{1}{SH_{z}(t)}\func{Re}\sum_{m,m^{\prime
}=-S}^{S}c_{m}^{\ast }c_{m^{\prime }}V_{mm^{\prime }}e^{i\Delta \Phi
_{mm^{\prime }}(t)},  \label{deltamzQuant}
\end{equation}
where $V_{mm^{\prime }}\equiv \left\langle m\left| V(\mathbf{S})\right|
m^{\prime }\right\rangle ,$ $\Delta \Phi _{mm^{\prime }}(t)\equiv \Phi
_{m}(t)-\Phi _{m^{\prime }}(t),$ and $\Phi _{m}(t)=-m\int^{t}dt^{\prime
}H_{z}(t^{\prime })+V_{mm}t.$ Dephasing stems from the nonequidistant
spectrum, i.e., from the nonlinear $m$-dependence of $V_{mm}.$ As
coefficients of the unperturbed state $c_{m}$ after crossing the resonance
are close to those of Eq. (\ref{Coherent}) and are localized around a
classical value $m_{0}=\cos \theta ,$ calculation in Eq. (\ref{deltamzQuant}%
) yields results similar to those for the toy model. Expanding $\delta
m_{z}(t)$ in $1/S$ similarly to that of Eq. (\ref{mplusminusExactSmallt})
yields a linear time divergence
\begin{equation}
\delta m_{z}(t)\varpropto tS^{-1}\cos \left[ \Phi _{H_{z}}(t)+\psi (t)\right]
,  \label{deltamzDiv}
\end{equation}
for the linear sweep $H_{z}(t)=vt,$ where $\Phi
_{H_{z}}(t)=\int^{t}dt^{\prime }H_{z}(t^{\prime })$ and $\psi (t)$ is a
slowly varying phase generated by the interaction: $\dot{\psi}(t)\sim v(%
\mathbf{m}(t)).$ This is exactly the raw result of DCE shown in Fig. \ref
{Fig-LZQ-PvsHz}b.

We have found the source for the divergence of the DCE and seen
that the divergent contribution $\delta m_{z}(t)$ in fact tends to
zero, if described properly by Eq. (\ref{deltamzQuant}) while the
QNS effect with $1/S$ corrections is entirely contained in the
asymptotic value $m_{z}(\infty )=m_{z}^{(\mathrm{As})}$ in Eq.
(\ref{DeltamzDef}). Now it is clear that one should simply project
out parts of $m_{z}(t)$ that are oscillating at large times. The
latter can be done, e.g., by the ``antiphasing''
\begin{equation}
\overline{m}_{z}(t)=\frac{1}{2}\left[ m_{z}\left( t+\frac{T_{Hz}}{4}\right)
+m_{z}\left( t-\frac{T_{Hz}}{4}\right) \right] ,  \label{Antiphase}
\end{equation}
where $T_{Hz}=2\pi /H_{z}(t)$ is the period of oscillations. This
transformation kills the diverging factor $t$ in Eq. (\ref{deltamzDiv}). To
improve convergence, one can repeat antiphasing several times, including
that with multiple precession frequencies arising due to nonlinearity. The
result shown in Fig. \ref{Fig-LZQ-PvsHz}b at $t\rightarrow \infty $ is in a
good accord with the exact numerical solution of the Schr\"{o}dinger
equation for $P\equiv P(\infty )$ that is shown by a horizontal line.

\begin{figure}[t]
\unitlength1cm
\begin{picture}(11,6)
\centerline{\psfig{file=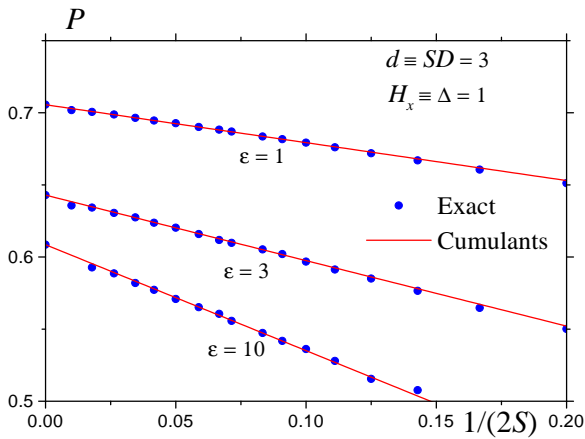,angle=-90,width=9cm}}
\end{picture}
\caption{ \label{Fig-LZQ-Pvs1S}
Quantum corrections to the final staying probability $P$
become more pronounced for slow sweep, $\varepsilon\gg 1$.
}
\end{figure}%
%

Fig. \ref{Fig-LZQ-Pvs1S} shows $P$ vs the quantum parameter $1/(2S)$ for
different sweep-rate parameters $\varepsilon $ of Eq. (\ref{PLZ}). The
agreement between the exact numerical solution and the result from DCE for $%
d\equiv SD=3$ (that corresponds to strong ferromagnetic interaction) and $%
H_{x}=1$ is very good. Particularly surprizing is the \emph{linearity} of $P$
vs $1/(2S)$ down to $S=5/2$! Quantum corrections to $P$ become more
pronounced for slow sweep, $\varepsilon \gg 1,$ and the slope of the curves
tends to infinity in the limit $\varepsilon \rightarrow \infty .$ This is in
accord with the finding of Ref. \cite{gar03prb}: For $h_{x}\equiv
H_{x}/(2d)<1$ in the classical case $\lim_{\varepsilon \rightarrow \infty
}P=(1-h_{x}^{2/3})^{3/2},$ whereas in the quantum case $\lim_{\varepsilon
\rightarrow \infty }P=0$ because of high-order tunneling analogous to
tunneling in molecular magnets.

We have seen above that divergence of DCE because of dephasing is in fact a
formal problem that can be solved to yield an excellent description of
quantum effects in the QNS model. However there is another more fundamental
problem related to the smoothness of the classical solution. For the
antiferromagnetic sign of the interaction, $d<0,$ the classical solution for
$P$ is a $\emph{nonmonotonic}$ function of $\varepsilon $ dropping fast to $%
P=0$ close to critical values $\varepsilon _{\nu },$ $\nu =1,2,\ldots $ (see
Figs. 6 and 8 of Ref. \cite{gar03prb}). $\varepsilon _{\nu }$ are
bifurcation points for classical trajectories at which for finite $S$ the
quantum distribution $\left| c_{m}(t)\right| ^{2}$ that was a narrow packet
at the beginning splits into two packets when crossing the resonance. One
can say that two parts of the quantum packet follow different classical
trajectories. In this case DCE cannot be expected to work well since the
assumption of a narrow packet close to a coherent state is not fulfilled.
Accordingly quantum corrections are very large, and one needs an extremely
large spin to approach the classical limit near these points (see Fig. 8 of
Ref. \cite{gar03prb}). The curves $P$ vs $1/(2S)$ are not straight in the
scale of Fig. \ref{Fig-LZQ-Pvs1S} and they are very different for different $%
\varepsilon .$

Classical models with more complicated interactions such as the biaxial
model with the transverse field along the hard axis show dynamical chaos for
$\emph{time}$ $\emph{dependent}$ $H_{z}$ that makes the classical nonlinear
spin switching effect poorly defined. In the chaotic regime the Schr\"{o}%
dinger equation provides a more adequate description of the problem than
classical equations of motion as it ``smoothes'' the chaos (see, e.g., Ref.
\cite{gut90}). As a result, quantum effects are much stronger than $1/S$
corrections in case of regular motion. On the other hand, for the field
along the medium axis classical motion is regular and smoothly depends on
parameters, thus DCE provides an excellent description of quantum
corrections at order $1/S$ that is comparable with that of Fig. \ref
{Fig-LZQ-Pvs1S}.

In this Letter, we have tested the dynamical cumulant expansion
for the particular model of interacting tunneling species with a
special interaction that is equivalent to a large spin. We have
shown that in the case of a smooth classical motion it works well.
This method can be generalized for systems with realistic
interactions that cannot be mapped onto a large spin and thus are
problematic if one uses the Schr\"{o}dinger equation for the whole
system. The system of cumulant equations will include that for the
magnetization $m_{\alpha ,i}$ that in the simplest case is
independent of the site $i$ and those for the pair cumulants
$m_{\alpha \beta ,ij}$ that depend on the distance between $i$ and
$j.$ One has to check however before applying DCE that the
classical (i.e., the mean-field) solution of the problem is not
chaotic.

\bibliographystyle{prsty}

\end{document}